\documentclass{elsart}

\usepackage[english]{babel}
\usepackage{amsmath,amsfonts,amssymb}
\usepackage{hyperref}
\usepackage[pdftex]{color,graphicx}

\begin{document}

\begin{frontmatter}

\title{On the equivalence of the label propagation method of community detection and a Potts model approach}

\author[BUTE]{Gergely Tib\'ely\corauthref{cor}},
\corauth[cor]{Corresponding author.}
\ead{tibelyg@maxwell.phy.bme.hu}
\author[BUTE,HAS]{J\'anos Kert\'esz}

\address[BUTE]{Department of Theoretical Physics, Budapest University of
Technology and Economics}
\address[HAS]{HAS-BME Solid State Physics Research Group\\
Budafoki \'ut 8, H-1111, Budapest, Hungary}

\begin{abstract}
We show that the recently introduced label propagation method for detecting communities in complex networks is equivalent to find the local minima of a simple Potts model. Applying to empirical data, the number of such local minima was found to be very high, much larger than the number of nodes in the graph. The aggregation method for combining information from more local minima shows a tendency to fragment the communities into very small pieces.
\end{abstract}


\end{frontmatter}

\section{Introduction}

The investigation of the community structure in complex networks is in the focus of network research  \cite{AB_rev}-\cite{Dor_book}. Special attention has been payed to the community structure of networks, i.e. the detection and analysis of groups of densely interconnected nodes \cite{HierOrgModMet}-\cite{labelprop}. The related problems range from the identification of functional modules in the biochemistry of the cell \cite{HierOrgModMet}, \cite{Amaral} to communities of people \cite{Newman_scitation}-\cite{socialgroupEv}, to name a few.

Recently, Raghavan \emph{et al.} suggested a method for detecting communities called \emph{label propagation} \cite{labelprop}. It defines a community as a set of nodes 
such that each node has at least as many neighbours in its own community as in any other one. In the initial stage of the method, all nodes form a distinct community (each node has an own ``label''). Then, at each timestep, the nodes join that community to which the largest fraction of their neighbours belong, by adopting the corresponding label. If there are multiple choices, a random decision is made with uniform distribution.   

In this note, we will show that the label propagation method is equivalent to a zero-temperature kinetic Potts-model, and investigate some of its properties on real-world graphs.

\section{The zero-temperature kinetic Potts model}

\begin{figure}[!h]
\begin{center}
\includegraphics*[height=1cm]{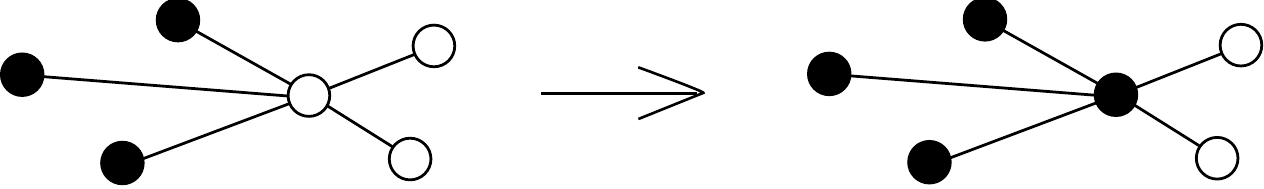}
\end{center}
\caption{Illustration of the dynamics: the center node has $3$ black and $2$ white neighbours, so it moves to the ``black'' state.}
\label{fig:ill} 
\end{figure}

Consider a $q$-state ferromagnetic Potts model on a graph $G$, placing spins on each vertex,
\begin{equation}  \label{eq:plain}
H(G)=-\sum_{ij}{A_{ij}\delta_{\sigma_i\sigma_j}}
\end{equation}
where $A_{ij}$ is the adjacency matrix: $A_{ij}=1$ if nodes $i$ and $j$ are connected, $0$ otherwise. The Potts variable $\sigma_i$ at node $i$ takes values from $1\ldots q$. We take $q=N$ where $N$ is the number of nodes in the network.
We define a zero-temperature kinetics in the following way. Lets start from a configuration where each spin is in a different state. Then, the spins are ordered into a random sequence. Following this sequence, each spin is aligned to the state, which contains the largest fraction of its neighbours; in case of ambiguity a random choice is made (see Fig. \ref{fig:ill}). 
After the last spin was considered, a new random sequence is established. The ground state is ferromagnetic, i.e., all spins are in the same state. However, the configuration may freeze into a metastable state, where more than one states are present. The latter depends on -- besides the adjacency matrix -- the random updating sequence and on the choices made in ambiguous cases. The criterium for the local minima is
\begin{equation}  \label{eq:crit}
k_i^{(\sigma_i)}\geq k_i^{(\sigma_j)}, \quad{\forall j,i}
\end{equation}
where $k_i^{(\sigma_j)}$ is the number of neighbours of node $i$ which are in the spin state of node  $j$ (denoted by $\sigma_j$). Finally, the communities are identified as sets of nodes in the same state. Note that this approach avoids the problematic issue of choosing a proper null-model. 

Clearly, this kinetic Potts model is the same as the one proposed by \cite{labelprop}. So the solutions of the label propagation method are the same as the local minima according to eq.\ref{eq:plain}.

We note that there is a related method, the superparamagnetic Potts model \cite{spc}, which uses a Hamiltonian similar to eq.\ref{eq:plain}. However, there the temperature should be chosen well above zero in the superparamagnetic regime. The correlations between spins are measured, and blocks in the correlation matrix are found by a simple thresholding procedure. In case of hierarchical organization or different average correlations in different communities, an elaborate investigation on different temperatures is needed.

It is straightforward to extend the method for weighted graphs, where the weights can be based on e.g. traffic/capacity, betweenness centrality \cite{betw}, Q-measures \cite{Qmes}, or other measures.

\section{Analysis of the method}

The zero-temperature kinetic model was investigated earlier on lattices \cite{freezing}, \cite{freezing2d3d}, \cite{freezing_potts}. It was reported that such dynamics may also end up in metastable states, instead of the ground state. In the $q=2$ (Ising) case on the  square lattice, the systems freezes in parallel stripes, with different alignment of spins in neighbouring stripes, with probability $p\approx1/3$ in 2D. On lattices with odd coordination number, finite frozen islands are also possible. In higher dimensions, the sytem will circulate forever over equienergetic configurations (``blinking'' states), with $p\to1$, as the number of spins diverges. Similar observations were made for $q>2$. So on regular graphs, the dynamics described in the previous section also leads to communities which are clearly irrelevant.
Although the high degree of symmetry makes the regular lattices very different from complex networks, regarding the possible metastable states, these findings indicate that it is important to check the relevance of the results, e.g. by comparing the outcomes of runs with different random number sequences.

We applied the kinetic Potts method described in the previous section to $2$ real-world graph, the Zachary karate club \cite{Zachary source} and the protein interaction network \cite{pin source}. As there may be many local minima, the algorithm was run several times on both datasets. In such situations, one should decide whether the actual local minimum was already found earlier. In order to save time, instead of comparing directly the two community sets, the quantity $\tilde{Q}=4L^2\cdot Q$ for the two configurations were compared, where $Q$ is the modularity \cite{Newman_basic}, and $L$ is the number of links in the network. If the $\tilde{Q}$ values were exactly the same, the two configurations were judged as being identical. Although improbable, it is theoretically possible that different configurations have the same modularity value. Thus the obtained numbers are lower bounds of the numbers of the visited different local minima.

The most prominent feature of the results is the proliferation of the local minima: the algorithm found $518$ different local minima (excluding the global one) out of $10^6$ runs for the karate club network (consisting of $34$ nodes), and $129691$ out of $2.2\times10^5$ runs for the PIN ($2111$ nodes). As the last new local minima were found in the $990767$th and in the $219999$th runs, respectively, most probably the limit has not been reached. 
\newline The phenomenon that usually there are more local minima was also noticed by \cite{labelprop}. However, the solutions they found were quite similar, therefore the fact that there are several of them was judged as unimportant. Moreover, it was not realized how large the number of the solutions is. However, such a huge number of possible solutions suggests an exhaustive exploration, which requires much more CPU time than the $\mathcal{O}($number of links$)$ time of a single run. The analysis of the results obviously requires a systematic and algorithmic investigation, which is not a straightforward task.   

Reference \cite{labelprop} also proposed an aggregation method for checking the similarity of different solutions. It consists of assigning the same `aggregated' label to nodes which appeared always in the same community in $n$ different solutions. Then a different aggregate should be made from $n$ another solutions. Finally, a similarity measure should be calculated between the two sets of aggregate labels. Such aggregates were reported to be quite similar, therefore using an aggregate instead of an arbitrarily chosen single solution was suggested. 

However, for the Zachary graph, combining the solutions gives a result very close to the initial state of the label propagation method, i.e. placing each node into a different community ($3$ communities of $2$ members and $1$ of $3$ remained). A similar tendency was observed on the PIN -- by aggregating all solutions, $94\%$ of the nodes belong to communities not bigger than $10$ nodes (and there is only one community bigger than $14$ nodes). \textcolor{red}{}So one should decide how many solutions to aggregate, which in turn determines the sizes of the communities, or the scale of the investigation, in other words, such that aggregating more solutions gives smaller communities.

\section{Conclusions}

We have shown that the label propagation method is equivalent to find the local energy minima of a simple zero-temperature kinetic Potts model. The number of local minima of two real-world networks turned out to be very large; it exceeds considerably the number of the nodes in the network. This feature suggests that the configurations corresponding to the local minima need further investigation. Aggregating the configurations tends to deconstruct the communities into smaller units, depending on the number of the aggregated configurations.

\section{Acknowledgement}

The authors wish to thank U.N. Raghavan for providing information about the label propagation algorithm, and F. Igl\'oi for providing references on the lattice model. The partial support of OTKA by grant K60456 is also acknowledged.

\end{document}